\documentclass[10pt,twocolumn,twoside]{IEEEtran}

\usepackage{tipa}
\usepackage{fixltx2e}
\usepackage{amssymb}
\usepackage{amsmath}
\usepackage{txfonts}
\usepackage{subfigure}
\usepackage{subfloat}
\usepackage{amssymb,algorithmic}
\usepackage{graphicx}
\usepackage[ansinew]{inputenc}
\input epsf
\usepackage{epic,eepic}
\usepackage{url}
\usepackage{algorithm}
\usepackage{color}
\usepackage{booktabs}
\usepackage{setspace}
\usepackage{cite}

\title{\LARGE Joint Transmit Diversity Optimization and Relay Selection for Cooperative MIMO Systems using Discrete Stochastic Algorithms
}

\author{\authorblockN{Patrick Clarke and Rodrigo C. de Lamare}\\
\authorblockA{Communications Research Group,
University of York,
York, YO10 5DD, UK.\\
Email: pjc500@ohm.york.ac.uk, rcdl500@ohm.york.ac.uk}
}

\begin{document}

\linespread{1.05}

\maketitle

\begin{abstract}
We propose a joint discrete stochastic optimization based transmit diversity selection (TDS) and relay selection (RS) algorithm for decode-and-forward (DF), cooperative MIMO systems with a non-negligible direct path. TDS and RS are performed jointly with continuous least squares channel estimation (CE), linear minimum mean square error (MMSE) receivers are used at all nodes and no inter-relay communication is required. The performance of the proposed scheme is evaluated via bit-error rate (BER) comparisons and diversity analysis, and is shown to converge to the optimum exhaustive solution.
\end{abstract}


\section{Introduction}
\label{Introduction}

\IEEEPARstart{C}{ooperative MIMO} networks have significant benefits in terms of diversity and robustness over non-cooperative networks. Consequently, they have been presented as a topology for the next generation of mobile networks \cite{Coop_comms_mobile_ad_hoc_networks_Scaglione}. Antenna selection, relay selection (RS) and diversity maximization are central themes in MIMO relaying literature \cite{Greedy_ant_selection_MIMO_relay_Ding, joint_source_relay_opt_MIMO_relay_Koshy, MIMO_antenna_discrete_optimization_Krishnamurthy}. However, current approaches are often limited to stationary, single relay systems and channels which assume the direct path from the source to the destination is negligible \cite{joint_source_relay_opt_MIMO_relay_Koshy}.

In this letter, the problems of transmit diversity selection (TDS) and RS are formulated as joint discrete optimization problems, where RS refines the set from which TDS is made; leading to improved convergence, performance and complexity. \textcolor{red}{Low-}complexity discrete stochastic algorithms (DSA) with mean square error (MSE) cost functions are employed to arrive at a solution. Continuous recursive least squares (RLS) channel estimation (CE) is introduced to form a combined framework, where adaptive RS and TDS are performed jointly with no forward channel state information (CSI). The proposed algorithms are implemented, and bit error-rate (BER) and diversity comparisons given against the exhaustive search solution and the unmodified cooperative system.

\section{System Model}
\label{sec:System_Model}

\begin{figure}[!t]
\centering
\includegraphics[width=0.75\columnwidth]{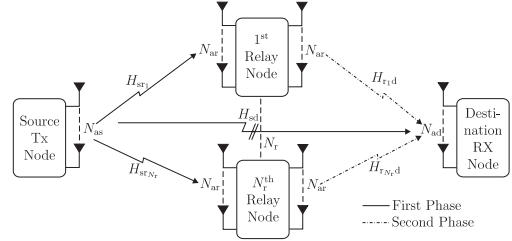}
\caption{MIMO multi-relay system model.}
\label{fig:System_model}
\end{figure}

We consider a QPSK, two-phase, decode-and-forward (DF), multi-relay MIMO system with half-duplex relays. Linear minimum mean square error (MMSE) receivers are used at all nodes and an error-free control channel is assumed \cite{Greedy_ant_selection_MIMO_relay_Ding, MIMO_antenna_discrete_optimization_Krishnamurthy}. All channels between antenna pairs are flat fading, have a coherence time equal to the period of an $N$ symbol packet and are represented by a complex gain. The direct path is non-negligible and has an expected gain of a fraction of that of the indirect paths; reflecting the increased distance and shadowing involved. An outline system model is given by Fig. \ref{fig:System_model}.
The system comprises $N_{\mathrm{r}}$ intermediate relay nodes which lie between single source and destination nodes which have $N_{\mathrm{as}}$ and $N_{\mathrm{ad}}$ antennas, respectively. The relay nodes have $N_{\mathrm{ar}}$ antennas, where $N_{\mathrm{ar}}$ is an integer multiple of $N_{\mathrm{as}}$ in order to reduce feedback requirements. The transmitted data consists of $N_{\mathrm{as}}$ independent \textcolor{red}{data streams}, \textcolor{red}{which are allocated to the correspondingly numbered antenna at the source and at the relays}. The source node transmits to the relay and destination nodes during the first phase, and the second phase involves the relay nodes decoding and forwarding their received signals to the destination. The maximum spatial multiplexing gain and diversity advantage simultaneously available in the system \textcolor{red}{are} $r^{*}=N_{\mathrm{as}}$ and $d^{*}=N_{\mathrm{ad}}(1+(N_{\mathrm{r}}N_{\mathrm{ar}}/N_{\mathrm{as}}))$, respectively \cite{diversity_and_multiplexing_tse}. The $N_{\mathrm{ad}}\times 1$ and $N_{\mathrm{ar}}\times 1$ first phase received signals at the destination and the $n^{\mathrm{th}}$ relay are given by
\begin{equation}
\mathbf{r}_{\mathrm{sd}}[i] = \mathbf{H}_{\mathrm{sd}}[i]A_{s}\mathbf{T}_{\mathrm{s}}\mathbf{s}[i] + \eta_{\mathrm{sd}}[i],
\end{equation}
\begin{equation}
\mathbf{r}_{\mathrm{sr}_{n}}[i] = \mathbf{H}_{\mathrm{sr}_{n}}[i]A_{\mathrm{s}}\mathbf{T}_{\mathrm{s}}\mathbf{s}[i] + \eta_{\mathrm{sr}_{n}}[i],
\label{eq:source_relay}
\end{equation}
respectively. The matrices $\mathbf{H}_{\mathrm{sd}}$ and $\mathbf{H}_{\mathrm{sr}_{n}}$ are the $N_{\mathrm{ad}}\times N_{\mathrm{as}}$ source - destination and $N_{\mathrm{ar}}\times N_{\mathrm{as}}$  source - $n^{\mathrm{th}}$ relay channel matrices, respectively. The subscripts s,d and $\mathrm{r}_{n}$ refer to the source, destination and $n^{\mathrm{th}}$ relay nodes, respectively. The quantity $\eta$ is a vector of zero mean additive white Gaussian noise, $\mathbf{s}$ is the $N_{\mathrm{as}}\times 1$ data vector, and $A_{\mathrm{s}}$ is the scalar transmit power allocation. The TDS matrix, $\mathbf{T}_{\mathrm{s}}$, is a $N_{\mathrm{as}}\times N_{\mathrm{as}}$ diagonal matrix where each element on the main diagonal specifies whether the correspondingly numbered antenna is active. The received signal of the second phase at the destination is the sum of the forwarded signals from the $N_{\mathrm{r}}$ relays and is expressed as
\begin{equation}
\mathbf{r}_{\mathrm{rd}}[i] = \boldsymbol{\boldsymbol{\mathcal{H}}}_{\mathrm{rd}}[i]A_{\mathrm{r}}\boldsymbol{\mathcal{T}}_{\mathrm{r}}[i]\mathbf{\hat{\bar{s}}}[i] + \eta_{\mathrm{rd}}[i],
\label{eq:rd_compound}
\end{equation}
where
$\boldsymbol{\mathcal{T}}_{\mathrm{r}}=\mbox{diag}\big[\boldsymbol{\mathbf{T}}_{\mathrm{r}_{1}} \ \boldsymbol{\mathbf{T}}_{\mathrm{r}_{2}}... \boldsymbol{\mathbf{T}}_{\mathrm{r}_{N_{\mathrm{r}}}} \big]$
is the $N_{\mathrm{ar}}N_{\mathrm{r}}\times N_{\mathrm{ar}}N_{\mathrm{r}}$ relay TDS matrix, $\mathbf{\hat{\bar{s}}}[i] = \big[\mathbf{\hat{s}}_{\mathrm{r}_{1}}^{T}[i]
$\begin{tiny}$
\overset{\frac{N_{\mathrm{ar}}}{N_{\mathrm{as}}}}
{\cdots}$\end{tiny}$ \mathbf{\hat{s}}_{\mathrm{r}_{1}}^{T}[i]
$\begin{tiny}$
\overset{}
{\cdots}$\end{tiny} $\mathbf{\hat{s}}_{\mathrm{r}_{N_{\mathrm{r}}}}^{T}[i]
$\begin{tiny}$
\overset{\frac{N_{\mathrm{ar}}}{N_{\mathrm{as}}}}
{\cdots}$\end{tiny}$ \mathbf{\hat{s}}_{\mathrm{r}_{N_{\mathrm{r}}}}^{T}[i]\big]$
 is the $N_{\mathrm{ar}}N_{\mathrm{r}}\times 1$ estimated data vector and $\boldsymbol{\boldsymbol{\mathcal{H}}}_{\mathrm{rd}}[i] = \big[\mathbf{H}_{\mathrm{r}_{1}\mathrm{d}}[i] \ \mathbf{H}_{\mathrm{r}_{2}\mathrm{d}}[i]... \mathbf{H}_{\mathrm{r}_{N_{\mathrm{r}}}\mathrm{d}}[i]\big]$ is the $N_{\mathrm{ad}}\times N_{\mathrm{ar}}N_{\mathrm{r}}$ channel matrix.

The Linear MMSE receiver at each relay is given by
\begin{equation}
\mathbf{W}_{\mathrm{sr}_{n}}[i]=\underset{\mathbf{W}_{\mathrm{sr}_{n}}}{\mbox{arg\,min}}\;E\Big[\big\Vert \mathbf{s}[i]-\mathbf{W}_{\mathrm{sr}_{n}}^{H}[i]\mathbf{r}_{\mathrm{sr}_{n}}[i]\big\Vert^{2}\Big],
\label{eq:relay_wiener_filter}
\end{equation}
resulting in the following Wiener filter,
$\mathbf{W}_{\mathrm{sr}_{n}} = \mathbf{R}_{\mathrm{sr}_{n}}^{-1}\mathbf{P}_{\mathrm{sr}_{n}}$, where
$\mathbf{R}_{\mathrm{sr}_{n}} = E\big[\mathbf{r}_{\mathrm{sr}_{n}}[i]\mathbf{r}_{\mathrm{sr}_{n}}^{H}[i]\big]$ and
$\mathbf{P}_{\mathrm{sr}_{n}} = E\big[\mathbf{r}_{\mathrm{sr}_{n}}[i]\mathbf{s}^{H}[i]\big]$ are the autocorrelation and cross-correlation matrices, respectively.
At the destination, the received signals are stacked to give $\mathbf{r}_{\mathrm{d}}[i] = \big[\mathbf{r}^{T}_{\mathrm{sd}}[i] \mathbf{r}^{T}_{\mathrm{rd}}[i]\big]^{T}$. The MMSE filter which operates upon $\mathbf{r}_{\mathrm{d}}[i]$ is given by
\begin{equation}
\mathbf{W}_{\mathrm{d}}[i]=\underset{\mathbf{W}_{\mathrm{d}}}{\mbox{arg\,min}}\;E\Big[\big\Vert \mathbf{s}[i]-\mathbf{W}_{\mathrm{d}}^{H}[i]\mathbf{r}_{\mathrm{d}}[i]\big\Vert^{2}\Big]
\label{eq:dest_wiener_filter}
\end{equation}
and the resulting Wiener filter is $\mathbf{W}_{\mathrm{d}}=\mathbf{R}_{\mathrm{d}}^{-1} \mathbf{P}_{\mathrm{d}}$ where $\mathbf{R}_{\mathrm{d}} = E\big[\mathbf{r}_{\mathrm{d}}[i]\mathbf{r}_{\mathrm{d}}^{H}[i]\big]$,
$\mathbf{P}_{\mathrm{d}} = E\big[\mathbf{r}_{\mathrm{d}}[i]\mathbf{s}^{H}[i]\big]$. A QPSK slicer follows MMSE reception at all nodes; the output of which is taken as the symbol estimate \cite{adapt_filt_Haykin}. Using (\ref{eq:relay_wiener_filter}) and (\ref{eq:dest_wiener_filter}), the MSE at the $n^{\mathrm{th}}$ relay and destination are given by $\sigmaup^{2}_{\mathbf{s}} - \mbox{trace}\big(\mathbf{P}_{\mathrm{sr}_{n}}\mathbf{R}^{-1}_{\mathrm{sr}_{n}}\mathbf{P}_{\mathrm{sr}_{n}}\big)$ and $\sigmaup^{2}_{\mathbf{s}} - \mbox{trace}\big(\mathbf{P}_{\mathrm{d}}\mathbf{R}^{-1}_{\mathrm{d,}}\mathbf{P}_{\mathrm{d}}\big)$, respectively, where $\sigmaup^{2}_{\mathbf{s}}=E\big[\mathbf{s}^{H}[i]\mathbf{s}[i]\big]$.

\section{Problem Statement}
\label{sec:problem_statement}

In this section, we formulate the joint TDS and RS task as a discrete combinatorial MSE problem. The TDS optimization problem is given by
\begin{multline}
\boldsymbol{\mathcal{T}}_{\mathrm{r}}^{opt} = \underset{\boldsymbol{\mathcal{T}}_{\mathrm{r}}\in \Omega_{\mathrm{T}}}{\arg\ \min}\ \mathcal{C}\big[i,\boldsymbol{\mathcal{T}}_{\mathrm{r}},\hat{\boldsymbol{\mathcal{H}}}_{\mathrm{rd}},\hat{\boldsymbol{\mathcal{H}}}_{\mathrm{sd}}\big]\\ = \underset{\boldsymbol{\mathcal{T}}_{\mathrm{r}}\in \Omega_{\mathrm{T}}}{\arg\ \min}\ E \Big[ \big\Vert\mathbf{s}[i]-\mathbf{W}_{\mathrm{d}}[i,\boldsymbol{\mathcal{T}}_{\mathrm{r}},\hat{\boldsymbol{\mathcal{H}}}_{\mathrm{rd}},\hat{\boldsymbol{\mathcal{H}}}_{\mathrm{sd}}]
\mathbf{r}_{\mathrm{d}}[i]\big\Vert^{2}\Big],
\label{eq:mmse_opt_function}
\end{multline}
where $\Omega_{\mathrm{T}}$ is the TDS matrix set of cardinality $|\Omega_{\mathrm{T}}|={{N_{\mathrm{ar}}N_{\mathrm{r}}}\choose{N_{\mathrm{asub}}}}$ and $N_{\mathrm{asub}}$ is the number of active relay antennas.

The performance and complexity of solutions to (\ref{eq:mmse_opt_function}) \textcolor{red}{depend on} $|\Omega_{\mathrm{T}}|$. We decrease $|\Omega_{\mathrm{T}}|$ whilst ensuring a minimum level of diversity by fixing the number of active antennas at $N_{\mathrm{asub}}<N_{\mathrm{ar}}N_{\mathrm{r}}$. However, $|\Omega_{\mathrm{T}}|$ is significant at modest levels of antennas and relays, e.g. $N_{\mathrm{r}}\geq4$ and $N_{\mathrm{as}}\geq2$. Further improvements can be achieved by a process we term RS. By removing one or more relays from consideration based on their MSE performance, the cardinality and quality $\Omega_{\mathrm{T}}$ is improved \textcolor{red}{without} restricting the \textcolor{red}{second-}phase channels available to the TDS process. TDS using this refined set then leads to the optimization of both phases.

The selection of the single highest MSE relay can be expressed as a discrete maximization problem given by
\begin{multline}
r_{n}^{opt} = \underset{r_{n} \in \Omega_{\mathrm{R}}}{\arg\ \max}\ \mathcal{F}\big[i,r_{n},\hat{\boldsymbol{\mathcal{H}}}_{\mathrm{sr}_{n}}\big]\\
=\underset{r_{n} \in \Omega_{\mathrm{R}}}{\arg\ \max}\ E\Big[\big\Vert\mathbf{s}[i]-\mathbf{W}_{\mathrm{sr}_{n}}^{H}[i,r_{n},\hat{\boldsymbol{\mathcal{H}}}_{\mathrm{sr}_{n}}] \mathbf{r}_{\mathrm{sr}_{n}}[i]\big\Vert^{2}\Big],
\label{eq:discrete_opt_RS}
\end{multline}
where $\Omega_{\mathrm{R}}$ is the set of candidate relays. Extension to the selection of multiple relays involves summing the MSE from candidate relays and populating $\Omega_{R}$ with sets of these relays. This results in $|\Omega_{\mathrm{R}}|={{N_{\mathrm{rem}}}\choose{N_{\mathrm{r}}}}$ where $N_{\mathrm{rem}}$ is the number of relays to be removed. Once RS optimization is complete, a refined subset, $\bar{\Omega}_{\mathrm{T}}\in \Omega_{\mathrm{T}}$, is generated by removing members of $\Omega_{\mathrm{T}}$ which involve transmission from $r_{n}^{opt}$. TDS then operates with this subset, where $|\bar{\Omega}_{\mathrm{T}}|={{N_{\mathrm{ar}}(N_{\mathrm{r}}-N_{\mathrm{rem}})}\choose{N_{\mathrm{asub}}}}$.

\section{Proposed Algorithm}
\label{sec:proposed_algorithms}

We propose a \textcolor{red}{low-}complexity DSA which jointly optimizes RS and TDS in accordance with (\ref{eq:mmse_opt_function}) and (\ref{eq:discrete_opt_RS}), and converges to the optimal exhaustive solution.
\begin{table}
\caption{Proposed discrete stochastic \textcolor{red}{joint TDS and RS}  algorithm}
\begin{spacing}{1}
\begin{footnotesize}
\begin{tabular}{l}
\toprule
\textbf{Step}\\
\textbf{1. Initialization}\\
\hspace{0.3cm}choose $r[1] \in \Omega_{\mathrm{R}}, r^{\mathrm{W}}[1] \in\Omega_{\mathrm{R}}$,
$\boldsymbol{\pi}_{\mathrm{R}}\big[1,r[1]\big]=1$,
$\boldsymbol{\pi}_{\mathrm{R}}[1,\tilde{r}]=0$ for $\tilde{r}\neq r[1]$\\
\textbf{2. For the time index} $i=1,2, ... , N$\\
\hspace{0.3cm}choose $r^{\mathrm{C}}[i] \in \Omega_{\mathrm{R}}$\\
\textbf{3. Comparison and update of the worst performing relay}\\
\hspace{0.3cm}if {$\mathcal{F}\big[i,r^{\mathrm{C}}[i]\big] > \mathcal{F}\big[i,r^{\mathrm{W}}[i]\big]$}
then $r^{\mathrm{W}}[i+1] = r^{\mathrm{C}}[i]$\\
\hspace{0.3cm}otherwise $r^{\mathrm{W}}[i+1] = r^{\mathrm{W}}[i]$\\
\textbf{4. State occupation probability (SOP) vector update}\\
\hspace{0.3cm}$\boldsymbol{\pi}_{\mathrm{R}}[i+1] = \boldsymbol{\pi}_{\mathrm{R}}[i] + \mu[i+1](\mathbf{v}_{r^{\mathrm{W}}[i+1]}-\boldsymbol{\pi}_{\mathrm{R}}[i])$ where $\mu[i] = 1/i$\\
\textbf{5. Determine largest SOP vector element and \textcolor{red}{select the optimum} relay}\\
\hspace{0.3cm}if {$\boldsymbol{\pi}_{\mathrm{R}}\big[i+1,r^{\mathrm{W}}[i+1]\big] > \boldsymbol{\pi}_{\mathrm{R}}[i+1,r[i]]$}
then $r[i+1] = r^{\mathrm{W}}[i+1]$\\
\hspace{0.3cm}otherwise $r[i+1] = r[i]$\\
\textbf{6. TDS Set Reduction}\\
\hspace{0.3cm}remove members of $\Omega_{\mathrm{T}}$  which utilize $r[i+1]$ ($\Omega_{\mathrm{T}} \rightarrow \bar{\Omega}_{\mathrm{T}})$\\
\bottomrule
\end{tabular}
\end{footnotesize}
\end{spacing}
\label{tab:RS_algorithm}
\end{table}
The RS portion of the DSA is given by the algorithm of Table \ref{tab:RS_algorithm}. At each iteration the MSE of a randomly chosen candidate relay ($r^{\mathrm{C}}$) (step 2) and that of the worst performing relay currently known ($\mathrm{r}^{\mathrm{W}}$) are calculated (step 3). Via a comparison, the higher MSE relay is designated $\mathrm{r}^{\mathrm{W}}$ for the next iteration (step 3). The current solution and the relay chosen for removal ($r$) is denoted as the current optimum and is the relay which has occupied $r^{\mathrm{W}}$ most frequently over the course of the packet up to the $i^{\mathrm{th}}$ time instant; effectively an average of the occupiers of $r^{\mathrm{W}}$. This averaging/selection process is performed by allocating each member of $\Omega_{\mathrm{R}}$ a $|\Omega_{\mathrm{R}}| \times 1$ unit vector, $\mathbf{v}_{l}$, which has a one in its corresponding position in $\Omega_{\mathrm{R}}$, i.e., $\mathbf{v}_{r^{\mathrm{W}}}[i]$ is the label of the worst performing relay at the $i^{\mathrm{th}}$ iteration. The current optimum is then chosen and tracked by means of a $| \Omega_{\mathrm{R}}| \times 1$ state occupation probability (SOP) vector, $\boldsymbol{\pi}_{\mathrm{R}}$. This vector is updated at each iteration by adding $\mathbf{v}_{r^{\mathrm{W}}}[i+i]$ and subtracting the previous value of $\boldsymbol{\pi}_{\mathrm{R}}$ (step 4).  The current optimum is then determined by selecting the largest element in $\boldsymbol{\pi}_{\mathrm{R}}$ and its corresponding entry in $\Omega_{\mathrm{R}}$ (step 5). Through this process, the current optimum converges towards and tracks the exhaustive solution \cite{global_search_discrete_opt_Andradottir}. An alternative interpretation of the proposed algorithm is to view the transitions, $r^{\mathrm{W}}[i]\rightarrow r^{\mathrm{W}}[i+1]$, as a Markov chain and the members of $\Omega_{\mathrm{R}}$ as the possible transition states. The current optimum can then be defined as the most visited state.

Once RS is complete at each time instant, set reduction ($\Omega_{\mathrm{T}}\rightarrow\bar{\Omega}_{\mathrm{T}}$, step 6) and TDS can take place. To perform TDS, modified versions of steps 1-5 are used. The considered set is replaced, $\Omega_{\mathrm{R}} \rightarrow \bar{\Omega}_{\mathrm{T}}$; the structure of interest is replaced, $r \rightarrow \boldsymbol{\mathcal{T}}_{\mathrm{r}}$; the best performing matrix is sought $r^{\mathrm{W}} \rightarrow \boldsymbol{\mathcal{T}}_{\mathrm{r}}^{\mathrm{B}}$; the SOP vector is replaced $\boldsymbol{\pi}_{\mathrm{R}}\rightarrow \boldsymbol{\pi}_{\mathrm{\boldsymbol{\mathcal{T}}}}$ and $\mathcal{C}\rightarrow \mathcal{F}$ from (\ref{eq:mmse_opt_function}). Finally, the inequality of step 3 is reversed to enable convergence to the lowest MSE TDS matrix which is the feedback to the relays.


Convergence of the proposed algorithm to the optimal exhaustive solution is dependent on the independence of the cost function observations and the satisfaction of $
\mathrm{Pr}\big\{\mathcal{F}\big[r^{\mathrm{opt}}[i]\big]>\mathcal{F}\big[r[i]\big]\big\} >\mathrm{Pr}\big\{\mathcal{F}\big[r[i]\big]>\mathcal{F}\big[r^{\mathrm{opt}}[i]\big]\big\}
$ and $
\mathrm{\mathrm{Pr}}\big\{\mathcal{F}\big[r^{\mathrm{opt}}[i]\big]>\mathcal{F}\big[r^{\mathrm{C}}[i]\big]\big\} >\mathrm{Pr}\big\{\mathcal{F}\big[r[i]\big]>\mathcal{F}\big[r^{\mathrm{C}}[i]\big]\big\}
$ for RS and TDS (with the afore mentioned modifications). In this work, to minimize complexity, independent observations are not used, therefore the proof of convergence is intractable. However, excellent convergence has been observed under these conditions in \cite{MIMO_antenna_discrete_optimization_Krishnamurthy} and throughout the simulations conducted for this work.

Significant complexity savings result from the proposed algorithm; savings which increase with $N_{\mathrm{as}}$, $N_{\mathrm{ar}}$, $N_{\mathrm{ad}}$, $N_{\mathrm{r}}$ and $N_{\mathrm{rem}}$. When $N_{\mathrm{r}} = 10$, $N_{\mathrm{as}}=N_{\mathrm{ar}}=N_{\mathrm{ad}}=2$, $N_{\mathrm{rem}}$ and $N_{\mathrm{asub}}=4$, the number of complex multiplications for MMSE reception and exhaustive TDS, exhaustive TDS with RS, iterative TDS and iterative TDS with RS are $5.8\times 10^{8}$, $1.7\times 10^{8}$, $1.8\times 10^{5}$ and $5.9\times 10^{4}$, respectively, for each time instant.

\section{Simulations}
\label{sec:Simulations}

In this section, simulations of the proposed algorithms (Iterative TDS with RS) are presented and comparisons drawn against the optimal exhaustive solutions (Exhaustive TDS with RS), the unmodified system (No TDS), and the direct transmission (Non-Cooperative). Plots of the schemes with TDS only (Exhaustive TDS, Iterative TDS) are also included to illustrate the performance improvement obtained by RS. Equal power allocation is maintained in each phase, where $A_{\mathrm{r}}=1/\sqrt{N_{\mathrm{asub}}}$ when TDS is employed and $A_{\mathrm{r}}=1/\sqrt{N_{\mathrm{ar}}N_{\mathrm{r}}}$ for the unmodified system. For the RLS CE, $\mathbf{P}_{\boldsymbol{\mathcal{\hat{H}}}_{\mathrm{rd}}}$, $\mathbf{P}_{\mathbf{\hat{H}}_{\mathrm{sr}_{n}}}$ and $\mathbf{P}_{\mathbf{\hat{H}}_{\mathrm{sd}}}$  are initialized as identity matrices and the exponential forgetting factor is 0.9. The initial values of $\boldsymbol{\mathcal{\hat{H}}}_{\mathrm{rd}}$, $\mathbf{\hat{H}}_{\mathrm{sr}_{n}}$ and $\mathbf{\hat{H}}_{\mathrm{sd}}$ are zeros matrices. Each simulation is averaged over 1000 packets ($N_{\mathrm{p}}$); each made up $N$ of pilot symbols.
\begin{figure}[!htb]
\begin{center}
\def\epsfsize#1#2{1\columnwidth}
\epsfbox{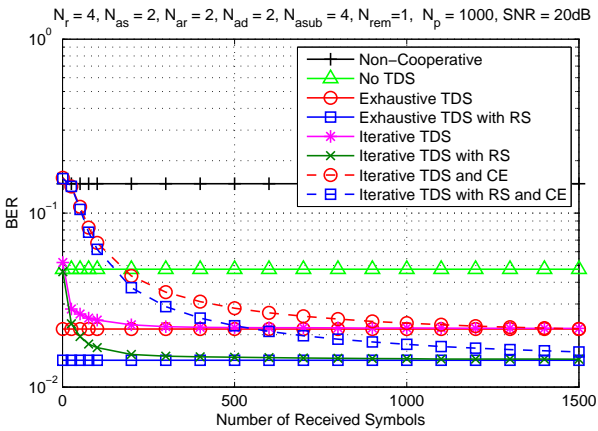} \caption{BER performance versus the number of
received symbols.} \label{fig:ber_ce}
\end{center}
\end{figure}
Fig. \ref{fig:ber_ce} gives the BER convergence performance of the proposed algorithms. The iterative TDS with RS algorithm converges to the optimal BER as does TDS with RS and CE, albeit in a delayed fashion due to the CE. The TDS with RS scheme exhibits quicker convergence and lower steady state BER. These results and the interdependence between elements of the algorithm confirm that both the RS and TDS portions of the algorithm converge to their exhaustive solutions but also the satisfaction of the probability conditions of Section \ref{sec:proposed_algorithms}.
\begin{figure}[!htb]
\begin{center}
\def\epsfsize#1#2{1\columnwidth}
\epsfbox{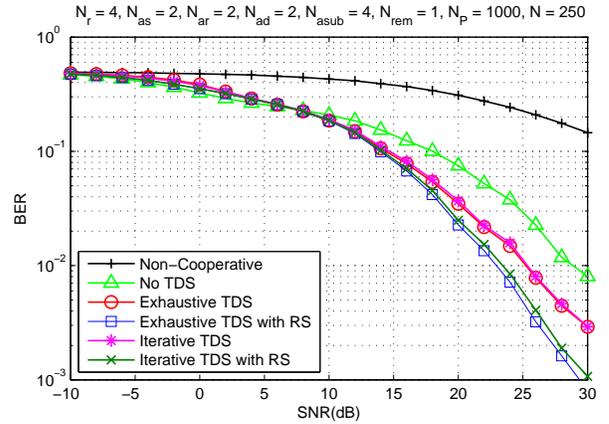} \caption{BER performance versus SNR.}
\label{fig:SNR}
\end{center}
\end{figure}

Fig. \ref{fig:SNR} shows the BER versus SNR performance of the proposed and conventional algorithms. Increased diversity has been achieved whilst maintaining $r^{\ast}$, illustrating that although the maximum available diversity advantage decreases with RS with TDS to  $d^{*}=N_{\mathrm{ad}} (N_{\mathrm{asub}}/N_{\mathrm{ar}}+1)$, the actual diversity achieved has increased. These diversity effects can be attributed to the removal of poor paths which bring little benefit in terms of diversity, but also the increase in transmit power over the remaining paths. The largest gains in diversity are present in \textcolor{red}{15dB-25dB} region and begin to diminish above this region because relay decoding becomes increasingly reliable and lower power paths become more viable for transmission.

\section{Conclusions}
\label{SEC:Conclusions}

This work presented a joint DSA which combines TDS and RS along with continuous CE for multi-relay cooperative MIMO systems.The scheme exceeds the performance of systems which lack TDS and matches that of the optimal exhaustive solution whilst saving considerable computational expense, making it ideal for realtime mobile use.

\end{document}